\documentclass[sigconf]{acmart}

\AtBeginDocument{%
  }

\setcopyright{none}
\acmConference{}{}{}
\acmBooktitle{}
\acmDOI{}
\acmISBN{}
\acmYear{}

\usepackage{multirow}
\usepackage{comment}
\usepackage{tcolorbox}
\usepackage{graphicx}
\usepackage{subcaption}
\usepackage{xcolor,soul,framed} 
\usepackage{multirow}
\usepackage{tikz,alltt}
\usepackage[table]{xcolor}
\usepackage{hyperref}
\usepackage{soul}

 \usetikzlibrary{arrows.meta,positioning,fit,calc,backgrounds}
\newcommand{\texttoplus}{\ensuremath{\oplus}} 

\newboolean{showcomments}
\setboolean{showcomments}{true} 

\ifthenelse{\boolean{showcomments}}{
	\newcommand{\nbc}[3]{
	{\colorbox{#3}{\bfseries\sffamily\scriptsize\textcolor{white}{#1}}}
	{\textcolor{#3}{\sf\small$\langle$\textit{#2}$\rangle$}}}
	}{
	\newcommand{\nbc}[3]{}
	
	}

\usepackage{tabularx}

\begin{document}
\title{Data-Dependent Goal Modeling for ML-Enabled Law Enforcement Systems}


\author{Dalal Alrajeh}
\affiliation{%
  \institution{Imperial College London, UK}
  \city{}
  \country{}
}
\email{dalal.alrajeh@imperial.ac.uk}

\author{Vesna Nowack}
\affiliation{%
  \institution{Imperial College London, UK}
  \city{}
  \country{}
}
\email{v.nowack@imperial.ac.uk}

\author{Patrick Benjamin\footnotemark[1]}
\affiliation{%
  \institution{University of Oxford, UK}
  \city{}
  \country{}
}
\email{patrick.benjamin@jesus.ox.ac.uk}

\author{Katie Thomas}
\affiliation{%
  \institution{University of Bath, UK}
  \city{}
  \country{}
}
\email{kt697@bath.ac.uk}

\author{William Hobson}
\affiliation{%
  \institution{University of Bath, UK}
  \city{}
  \country{}
}
\email{wghobson1@gmail.com}

\author{Carolina Gutierrez Muñoz}
\affiliation{%
  \institution{University of Bath, UK}
  \city{}
  \country{}
}
\email{cigm20@bath.ac.uk}

\author{Catherine Hamilton-Giachritsis}
\affiliation{%
  \institution{University of Bath, UK}
  \city{}
  \country{}
}
\email{chg26@bath.ac.uk}

\author{Juliane A. Kloess\footnotemark[2]}
\affiliation{%
  \institution{University of Glasgow, UK}
  \city{}
  \country{}
}
\email{juliane.kloess@glasgow.ac.uk}

\author{Jessica Woodhams}
\affiliation{%
  \institution{University of Birmingham, UK}
  \city{}
  \country{}
}
\email{j.woodhams@bham.ac.uk}

\author{Daniel Butler}
\affiliation{%
  \institution{Independent researcher, UK}
  \city{}
  \country{}
}
\email{db14921066@gmail.com}

\author{Mark Law}
\affiliation{%
  \institution{ILASP, UK}
  \city{}
  \country{}
}
\email{mark@ilasp.com}

\author{Ralph Morton}
\affiliation{%
  \institution{Aston University, UK}
  \city{}
  \country{}
}
\email{r.morton2@aston.ac.uk}

\author{Benjamin Costello}
\affiliation{%
  \institution{University of Birmingham, UK}
  \city{}
  \country{}
}
\email{b.d.costello@bham.ac.uk}

\author{Amy Burrell}
\affiliation{%
  \institution{University of Birmingham, UK}
  \city{}
  \country{}
}
\email{a.burrell@bham.ac.uk}
 
\author{Tim Grant}
\affiliation{%
  \institution{Aston University, UK}
  \city{}
  \country{}
}
\email{t.d.grant@aston.ac.uk}

\author{Prachiben Shah}
\affiliation{%
  \institution{University of Birmingham, UK}
  \city{}
  \country{}
}
\email{pxs907@student.bham.ac.uk}

\author{Frances Laureano de Leon}
\affiliation{%
  \institution{University of Birmingham, UK}
  \city{}
  \country{}
}
\email{fxl846@student.bham.ac.uk}

\author{Mark Lee}
\affiliation{%
  \institution{University of Birmingham, UK}
  \city{}
  \country{}
}
\email{m.g.lee@bham.ac.uk}

\renewcommand{\shortauthors}{Alrajeh et al.}
\begin{abstract}

\noindent
Investigating serious crimes is inherently complex and resource-constrained. Law enforcement agencies (LEAs) grapple with overwhelming volumes of offender and incident data, making effective suspect identification difficult. Although machine learning (ML)-enabled systems have been explored to support LEAs, several have failed in practice. This highlights the need to align system behavior with stakeholder goals early in development, motivating the use of Goal-Oriented Requirements Engineering (GORE).

This paper reports our experience applying the GORE framework KAOS to designing an ML-enabled system for identifying suspects in online child sexual abuse. We describe how KAOS supported early requirements elaboration, including goal refinement, object modeling, agent assignment, and operationalization. A key finding is the central role of data elicitation: data requirements constrain refinement choices and candidate agents while influencing how goals are linked, operationalized, and satisfied. Conversely, goal elaboration and agent assignment shape data quality expectations and collection needs.

Our experience highlights the iterative, bidirectional dependencies between goals, data, and ML performance. We contribute a reference model for integrating GORE with data-driven system development, and identify gaps in KAOS, particularly the need for explicit support for data elicitation and quality management. These insights inform future extensions of KAOS and, more broadly, the application of formal GORE methods to ML-enabled systems for high-stakes societal contexts.

\end{abstract}




\maketitle

\footnotetext[1]{Patrick Benjamin was at Imperial College London when this work was undertaken.}
\footnotetext[2]{Juliane A. Kloess was at University of Birmingham and University of Edinburgh when this work was undertaken.}



\section{Introduction}

Investigating serious crimes is an inherently complex and resource-constrained challenge. Law enforcement agencies (LEAs) must analyze overwhelming volumes of offender and incident data, and with limited investigative capacity. A key task is {suspect identification}, where the goal is to identify persons of interest likely to be involved in cases of criminal activity. However, the scale and heterogeneity of available data make this task extremely difficult to perform reliably without computational support. Furthermore, for crimes like child sexual abuse, the rapid advancements of the digital world and technology also mean that children have access to smartphones, the internet, social media and gaming platforms, increasing their vulnerability to experiencing exploitation and abuse~\cite{NCA_2025_CSA_Threat}.

Machine learning (ML)-enabled systems have been widely explored to augment investigative processes by detecting patterns, filtering large datasets, and providing ranked lists of suspects \cite{Nalchigar2021,JAVAID202258,AHMED2024101933,Sarzaeim2023SysReview}. Despite technical advances, several ML-enabled systems have failed to gain adoption or deliver consistent value in practice, e.g., \cite{griffard2019patternizr,scl2020santacruz,house2020rekognition,macarthur2021shotspotter}. A central reason is misalignment between system behaviour and stakeholder goals: predictive accuracy alone does not guarantee outputs are actionable, trustworthy, or aligned with legal and ethical constraints \cite{SANTOW2024}. High-profile cases, such as wrongful arrests due to misused facial recognition \cite{ACLU_2020_Williams_arrest}, highlight how inadequate alignment between technical design and stakeholder intent can have profound consequences in high-stake domains.


This motivates the need for requirements engineering approaches that explicitly connect system functionality, human roles, and organizational goals. 
Among existing approaches, Goal-Oriented Requirements Engineering (GORE) has been identified as particularly well-suited for structuring stakeholder intent, reasoning about trade-offs, and supporting the management of evolving requirements \cite{Lamsweerde2009,Yu1997,Horkoff2016}. GORE’s emphasis on decomposing high-level objectives into operationalizable requirements is especially relevant for ML-enabled systems, where data availability, model performance, and stakeholder expectations are tightly interdependent \cite{batista2024teaching}. However, while prior research has argued for the promise of goal-based methods in AI system design, there remains limited evidence of their practical application in real-world ML contexts \cite{Mavin2017,batista2024teaching}.

\noindent
\textbf{Aim and scope.} This paper addresses this gap by presenting an experience report on applying the GORE framework KAOS \cite{Lamsweerde2009} to the design of an ML-enabled, decision-support system within an LEA. The system supports the prioritization of suspects in cases of online sexual offending against children, a domain where the stakes for investigative accuracy and efficiency are particularly high.  A deliberate reason for choosing KAOS over other GORE frameworks is its strong formal foundations. KAOS offers precise semantics for goals, agents, objects, and refinements, enabling rigorous reasoning about satisfaction, conflicts, and responsibilities. By applying KAOS in this setting, we aim to expose the specific challenges that arise when formal RE frameworks are confronted with data-driven applications.

\noindent
\textbf{Contributions.} The contributions of this paper are threefold:
\begin{itemize}
    \item We present one of the first detailed accounts of applying GORE in the development of an ML-enabled system in a law enforcement setting, reporting on the challenges and adaptations required in practice;
    \item We provide a reference model for integrating GORE with data-centric ML, facilitating traceability between the requirement and implementation in ML-enabled software development;     \item We identify key lessons and open research gaps in applying KAOS to ML-enabled systems, highlighting how structured goal models can improve alignment between stakeholder intent, data artefacts, and implementation decisions. 
\end{itemize}

Together, these contributions demonstrate how GORE can act as a practical bridge between stakeholder intent and the design of accountable, trustworthy ML-enabled decision-support systems in high-stakes investigative contexts.

\section{Related Work}
\label{section:related}

\noindent
\textbf{Experience Reports on ML Development.}
Several studies have examined the practical challenges of developing ML-enabled systems. 
For example, Pei et al.~\cite{Pei2022} explored collaboration patterns among domain experts, software engineers, and data scientists in data-driven intelligent system development. 
Binder et al.~\cite{Binder2021} investigated the benefits of integrating user research into AI development processes, highlighting its impact on AI acceptance. 
Other work has reported empirical insights from industry: Vogelsang et al.~\cite{Vogelsang2019} studied how data scientists elicit and analyze requirements for ML systems, while Amershi et al.~\cite{Amershi2019} analyzed Microsoft teams developing customer-facing AI features, identifying best practices and process maturity indicators. 
These studies provide valuable lessons on ML development in practice, but they do not explicitly employ goal-oriented frameworks for requirements engineering.
A systematic literature review by Nahar et al.~\cite{Nahar2023} synthesized challenges faced by practitioners when building ML-enabled systems, highlighting, among others, the difficulty of mapping high-level business goals to concrete ML requirements. They also reported that testing and debugging ML models remains difficult due to the lack of clear specifications, making it challenging to define acceptable quality criteria, curate representative test data, and evaluate model robustness. Our paper contributes to this body of work by reporting on our experience in addressing these challenges through the application of GORE with the KAOS framework.

\noindent
\textbf{Requirements Engineering for ML Systems.}
A growing body of research examines the intersection of RE and ML.
Villamizar et al.~\cite{Villamizar2021} conducted a systematic literature review of RE for ML, identifying subject areas and quality attributes, and pointing to a gap in practical validation.
Other work has analyzed how software engineering practices differ for ML versus non-ML systems \cite{Wan2021HowDoesMLChange,Martinez2022}, finding that ML performance often depends critically on data availability.
Heyn et al.~\cite{heyn2021requirement} and Iqbal et al.~\cite{Iqbal2018} examined how uncertainty and probability in ML outputs affect requirements specification.
Together, these highlight unique RE challenges for ML but offers limited insights on formal methods like GORE.

Goal-oriented methods have been proposed as particularly promising for ML-enabled systems. 
Barrera et al.~\cite{barrera2024extension} extended goal modeling languages to capture ML-specific requirements, such as evaluation metrics and model types. 
Batista et al.~\cite{batista2024teaching} showed that KAOS can be effectively taught and applied to AI systems, although their study remained limited to requirements elicitation. 
Work by Ishikawa et al.~\cite{ishikawa2020evidence} introduced the notion of \emph{feasibility} to handle the uncertainty of goals that depend on ML model performance. 
More generally, researchers have extended goal models to incorporate probabilistic reasoning about requirements satisfaction, e.g., Liaskos et al.~\cite{Liaskos2022ModelingUncertainty}, Baslyman et al.~\cite{Baslyman2022Reasoning}, and Letier and Van Lamsweerde \cite{Letier2004Reasoning}. 
Our work builds on this line of research by applying a formal GORE framework KAOS in practice, and by explicitly linking goal models with data availability and ML evaluation outcomes.

Complementary to our work, the $M^3S$ approach~\cite{Husen2024} proposes and evaluates an integrated modeling and tool-supported framework to facilitate a consistent and comprehensive analysis of ML systems. $M^3S$ combines KAOS goal models with views such as ML Canvas and Architecture models, enabling the decomposition of business objectives into measurable ML performance requirements, which can then be validated against model outputs. While this work focuses on proposing and assessing a modeling framework, our contribution differs by providing an experience report that illustrates the practical use of KAOS and its potential extensions in the policing domain.
The GR4ML framework was empirically evaluated in~\cite{Nalchigar2021}  for its expressiveness and usefulness in supporting requirements elicitation and design of ML solutions. It provides three complementary modeling views that mediate between business stakeholders, data scientists, and data engineers. In parallel, RM4ML~\cite{Yang2024} instead proposes a UML-based requirements model for ML-enabled systems, comprising five diagrams that capture functional, quality, environmental, and data requirements. RM4ML emphasizes data aspects, such as content, quality, and volume, and extends class diagrams to incorporate environment classes. Unlike these works, which primarily propose new modeling languages or frameworks, our work contributes an experience-based account of applying a formal GORE method (KAOS) in practice, demonstrating its utility and exposing systematic gaps when requirements depend on empirical data and probabilistic ML performance. 

Finally, \cite{horkoff2019non} examined how non-functional requirements (NFRs) apply to ML-based systems, focusing on qualities tied to key functional requirements such as accuracy. They argue that traditional methods for specifying, refining, and reasoning about NFRs require adaptation or new techniques.

\noindent
\textbf{Data-Driven Requirements Elicitation}
There is a parallel stream of research on data-driven requirements elicitation. Maalej et al.~\cite{Maalej2016} introduced \textit{data-driven requirements engineering}, advocating the use of runtime feedback and large-scale usage data to guide requirements decisions beyond traditional stakeholder elicitation. Later, Maalej~\cite{Maalej2019} highlighted how explicit feedback (e.g., app store reviews, social media) and implicit feedback (e.g., usage data, error logs, sensor data) enable user-centered identification, prioritization, and management of requirements, allowing developers to aggregate and mine such data to inform release planning and selection. Surveys by Zowghi et al.~\cite{Zowghi2005} and Lim et al.~\cite{Lim2021} review approaches to deriving requirements from diverse data sources such as event logs and social media.  
In contrast, the data in our project does not express requirements directly but is crucial to their \emph{operationalization}. Such data-driven approaches could, however, support refining and evolving requirements once the tool is deployed, based on real-world usage and feedback.

\section{Goal-oriented modeling with KAOS}
\label{section:preliminaries}

Over the past decades, several goal modeling languages and approaches have been developed for expressing system goals and their interrelations. Prominent examples include KAOS \cite{Lamsweerde2009}, $i^*$ \cite{dalpiaz2016istar20languageguide}, Tropos \cite{morandini2014tropos}, and the NFR framework \cite{chung2012non}. Among these, we focus on KAOS, as it provides a clear separation between high-level strategic goals, operational requirements, and assumptions about the deployment environment. This separation is particularly important in ML-enabled systems, where system behaviour depends not only on software but also on human and organizational actors, data availability, and regulatory constraints. 

KAOS (Knowledge Acquisition in autOmated Specification) is a goal-oriented requirements engineering method that captures stakeholder intent through goals, their refinements, and the responsibilities of system components. A \textit{goal} is a prescriptive statement of intent to be satisfied by cooperating agents. The \textit{system} includes both software and its environment — people, legacy software, and devices such as sensors and actuators. An \textit{agent} is an active system component with specific \textit{responsibilities} for goal satisfaction and \textit{capabilities} defined by the conditions it can monitor or control. A \textit{domain property} is a descriptive statement about the system’s operational environment, such as a  regulatory constraint.

A \textit{goal model} in KAOS is expressed as an AND/OR refinement graph. 
 An \textit{AND-refinement} decomposes a parent goal into subgoals and domain conditions that together ensure its satisfaction. 
    An \textit{OR-refinement} represents alternative decompositions for achieving the same goal. 
     \textit{Leaf goals} are assigned to single agents. If assigned to a software agent, the goal becomes a \textit{software requirement}; if assigned to an environment agent, it is treated as an \textit{environment assumption}. 
In diagrammatic representations, goals are typically shown as parallelograms, domain properties as trapezoids, and agents as hexagons. Refinement links are captured by edges connected through a white circle, while directed edges from agents denote responsibility assignments.

\begin{figure}[h]
    \center
    \includegraphics[width=0.45\textwidth]{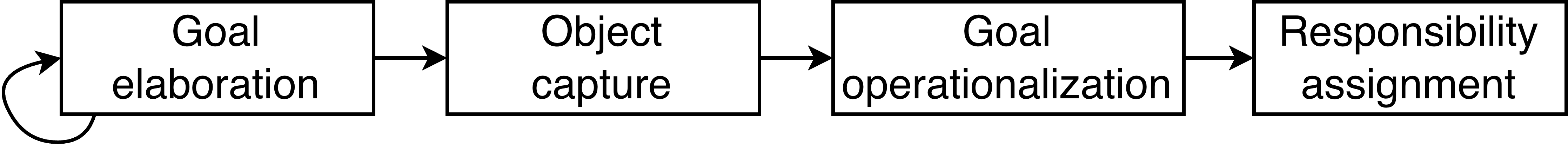}
    \caption{Traditional KAOS-based methodology.}
    \label{goal_diagram}
\end{figure}

KAOS complements goal models with an \textit{object model}, which captures the key concepts, entities, and relationships in the problem domain. This provides the semantic grounding necessary to make goals precise and verifiable. 
Goals are then \textit{assigned} to agents---human or software---who are responsible for their satisfaction. \textit{Agent assignment} bridges the gap between high-level stakeholder intent and concrete operational responsibilities. 

\textit{Goal operationalization} is the process of refining abstract goals into concrete operations that agents can perform. In KAOS, this involves progressively decomposing high-level objectives until the leaf goals are sufficiently precise to be assigned to specific agents or implemented as system requirements.  
For software-controlled goals, operationalization typically leads to the specification of operations constrained by domain properties and assumptions. 
Fig.~\ref{goal_diagram} summarizes the four main KAOS activities: \textit{goal elaboration}, \textit{object capture}, \textit{responsibility assignment}, and \textit{goal operationalization}~\cite{Lamsweerde00}.

\section{Project Description}
\label{section:project_description}

This study is part of an ongoing collaboration, initiated in late 2020, between four academic institutions and a LEA in the United Kingdom. The project focuses on the design of an ML-enabled decision-support system to assist investigators in identifying cases of online sexual offending against children. This domain presents a pressing challenge: investigators face overwhelming volumes of offender and incident data, and decisions about which cases to prioritise have direct consequences for the safety of potential victims. The central objective of the collaboration is therefore to explore how AI can support identification tasks while ensuring alignment with stakeholder goals, organizational constraints, and ethical principles.  

\subsection{Collaboration Structure and Roles}
The academic consortium brought together expertise in software engineering, natural language processing, machine learning, forensic linguistics, and forensic and clinical psychology. All academic partners obtained the necessary security clearance, and all project activities were conducted following being granted full ethical approval across all academic institutions.
The division of responsibilities within the academic team was as follows:
\begin{itemize}
    \item \textbf{Software Engineers (SE):} four experts elicited, modeled and refined requirements, performed data cleansing and pre-processing, developed the full pipeline.
    \item \textbf{Forensic and Clinical Psychologists  (FCP):} nine experts led stakeholder interviews, identified behavioural features and contributed to behavioural coding of the data. 
    \item \textbf{Forensic Linguists  (FL):} two experts identified linguistic features and contributed to linguistic coding of the data.
    \item \textbf{Machine Learning developers (ML):} four experts developed ML models using the annotated datasets to explore predictive capabilities aligned with stakeholder needs.
\end{itemize}

\subsection{Datasets and Preparation}


The LEA provided access to four datasets over the course of the project. Each dataset contained multiple cases, and each case included one or multiple files of textual data. The first (D1) and second (D2) datasets were unstructured textual data from Dark Web communities (n = 198 cases). The third (D3) and fourth (D4) datasets were unstructured textual data from Surface Web communities (n = 141 cases). D1 contained conversational data derived from transcripts of chat logs, forum posts, and private/instant messages. D2 included conversational data derived from forum posts and private/instant messages. D3 comprised conversational data from group chats on various messaging apps, and D4 consisted conversational data from a public chat room.
This heterogeneity posed specific challenges for requirements specification, labelling practice, and ML model training (see Section \ref{section:applying_KAOS}).

{\footnotesize
 \begin{table}[htbp]
    \caption{Dataset descriptions.}
    \centering
    \begin{tabular}{c c c c c c c c}
        \toprule
        \multirow{2}{*}{Datasets} & \multirow{2}{*}{\#cases}  &  \multirow{2}{*}{\#files} & \multirow{2}{*}{\#pages} &  \multicolumn{2}{c}{\#coded examples} \\
        \cline{5-6}
        & & & & behavioural & linguistic \\
        \hline
        D1 & 53 & 1-12  & 825 & 13,198 & 26,621 \\
        D2 & 145 & 1-2  & 3,766 & 25,100 & 42,424   \\
        D3 & 8 & 1-19  & 54 & 2,732 & 5,819 \\
        D4 & 133 & 1  & 444 & 47,946 & 36,058  \\
        \midrule
        Total & 339 & - & 5,089 & 88,976 & 110,922  \\
        \bottomrule
        \end{tabular}
        \label{table:data}
        \end{table} 
}

All datasets were redacted by the LEA before release to remove any personally identifying information and child sexual abuse material. Additional redaction of potentially sensitive information was carried out by forensic psychology and linguistics team members. The datasets were then reformatted and imported into NVivo\footnote{\url{https://lumivero.com/products/nvivo-qualitative-data-analysis-software/}}, a qualitative data analysis tool used for coding and annotation.
Table~\ref{table:data} summarizes the datasets, including their scale and the number of coded examples produced for behavioural and linguistic analyses, which will be discussed in the next section.

\section{Applying KAOS in Practice}
\label{section:applying_KAOS}

We report here on our experience of how a KAOS-driven approach guided our understanding of requirements for an ML-enabled decision-making system, the datasets available for decision-making, and the development of the system itself. 
Following the design science research paradigm \cite{Hevner2010}, we iteratively combined stakeholder and data-driven insights to guide our modelling and system development activities.

In our setting, however, the traditional structure of KAOS processes proved insufficient because progress in refining goals and assigning responsibilities was tightly constrained by the availability, quality, and labelling of datasets. To address this gap, we extended KAOS with an explicit \textit{data elicitation process}. We also introduced new interactions between this process and the traditional KAOS processes, as well as additional interactions among the traditional processes themselves.

\begin{figure*}
\center
\includegraphics[width=0.8\textwidth]{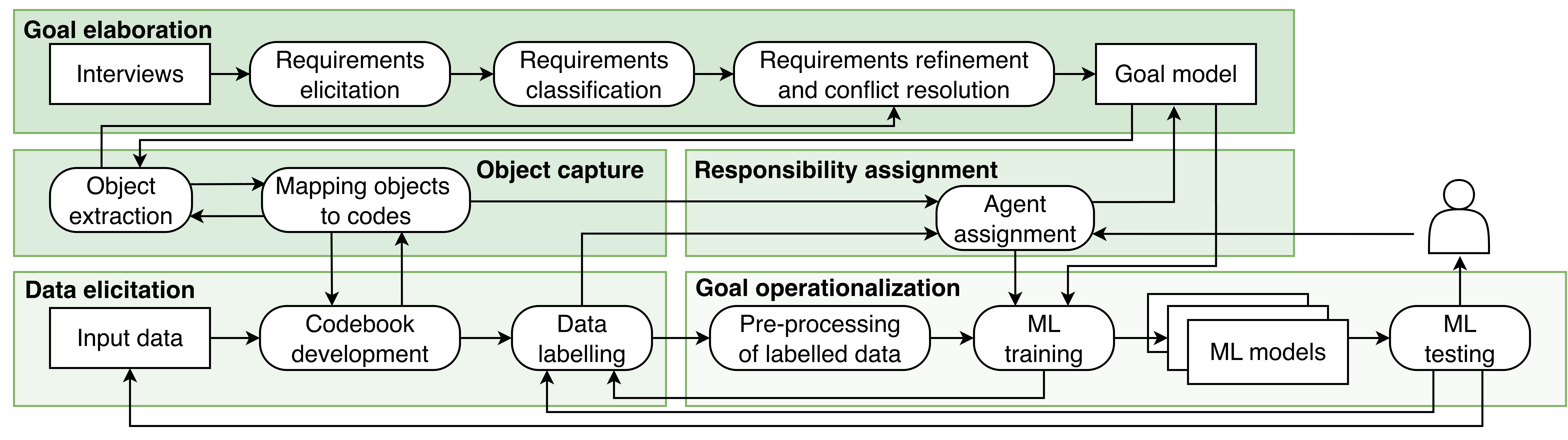}
\caption{The reference model we followed in developing the ML-enabled system.}
\label{figure:framework}
\end{figure*}

Fig.~\ref{figure:framework} summarises the resulting processes and their interactions. We performed goal elaboration by eliciting requirements from stakeholders and structuring them into a KAOS goal model to support subsequent responsibility assignment. This activity was conducted in parallel with data elicitation, which focused on identifying relevant code artefacts and data sources, assessing dataset availability and quality, and producing labelled data for ML. The two activities were linked through object modelling, where domain and goal concepts were captured and mapped to concrete data representations such as labels. 
We then carried out goal operationalization by translating selected goals into ML tasks. 
Achieved performance, constrained by the outcomes of data elicitation, directly informed responsibility assignment to ML agents and, where necessary, triggered revisions to goals, agent assignments or data.


\subsection{Goal elaboration}

This section focuses on the goal elicitation and refinement processes, highlighting how KAOS shaped our activities and the adaptations required in practice. 

\noindent
\subsubsection{Context.}
We began by eliciting stakeholder goals and domain properties through a series of semi-structured interviews with potential users (crime investigators from relevant teams). Semi-structured interviews were chosen because they allowed flexibility to probe into follow-up topics and provided participants with control over the issues discussed \cite{William2015}. Each interview covered both current practices (the \textit{as-is} system) and needs for a future ML-enabled system (the \textit{to-be} system). Due to the confidential nature of the research project, we are not able to share the interview questions related to current practices, but the questions related to the ML-enabled system can be found here: \href{https://github.com/vesnowack/Interview-questions/blob/main/Qs.pdf}{list of questions}.

Participants were recruited by circulating an invitation email to all members of four investigative teams within the law enforcement agency. Interested staff contacted the project liaison officer. Eleven investigators participated; some also held team manager roles but continued practical investigative work. For confidentiality, participants were identified only by first name. Most interviews were conducted in person, with a few held securely via teleconference or written responses. Due to security restrictions, only summary notes (not verbatim transcripts) were taken.

The summary notes were then qualitatively analysed using a Thematic Analysis approach \cite{Clarke2014} by four FCPs. This involved identifying themes and sub-themes that characterized current investigative practices and participants' views on the potential use of tool-based approaches. To ensure the validity of the findings, a summary of the emerging themes, particularly those describing the different team roles, was shared and discussed with two team managers to check whether the descriptions accurately reflected expected practice.

An initial set of key goals was identified and presented to higher-level managers for refinement and prioritization. A RAG rating exercise (red/amber/green) distinguished goals of varying importance. To strengthen reliability, two researchers (FCP and SE) independently derived additional goals from interview notes, focusing on missing refinements or assumptions needed to realize higher-priority goals. Their results were then compared in moderation sessions with a software engineering expert. Disagreements centred on project scope, implied environment assumptions, and fidelity to stakeholder intent.

When analysing notes, the researchers spent considerable time discussing terminology, identifying synonyms
and deciding on the format of requirements, leading to 'Moderate' agreements according to Fleiss' Kappa \cite{fleiss1971measuring} agreement scores (see Table \ref{table:interraterInterviews}). After that, the researchers became more confident and aligned with each other
for the remaining interviews, having `Substantial'/`Almost perfect' agreements for the rest of the interview notes. Notes from a joint interview with two participants (marked as 12* in the table) were analysed for both participants together.

\begin{figure}[t]
  \includegraphics[width=0.47\textwidth]{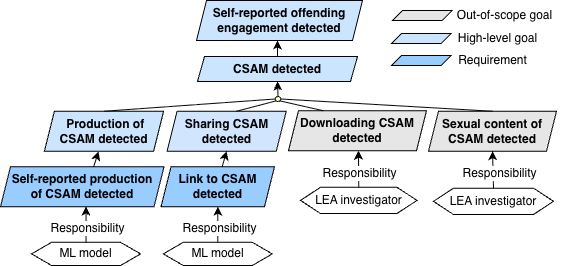}
  \caption{A partial goal model for the LEA system.} 
  \label{figure:KAOS}
  \end{figure}

In total, 429 goals matched between researchers, and 474 required discussion. After removing duplicates, 557 unique goals were formulated and classified into functional (499) and non-functional (58). The functional goals informed a provisional goal model (some subsumed others, others were refinements). From these, 203 environment assumptions were identified. The remaining 296 goals were further discussed, and 30 software requirements were deemed within scope and became the focus of prototype development. Fig.~\ref{figure:KAOS} shows a fragment of the derived goal model that refers to the detection of production of Child Sexual Abuse Material (CSAM).

Several requirements elicited from stakeholders were explicitly AI-specific upon prompting. For example, participants expressed a willingness to contribute feedback to help the system improve its predictions over time, highlighting the assumption that investigators would take on a new responsibility in supporting model learning (e.g., ``happy to spend time providing feedback to help improve the tool’s future use'' (P1)). At the same time, they stressed that human expertise and intuition would remain indispensable, noting that there are cases where ``a tool would be helpful'' (P5), but others where only human judgment could apply. Finally, participants requested that the system generate  justifications for its outputs, as a way to calibrate trust in the tool and provide transparency into its reasoning (e.g., ``a confidence score might give them more trust in the tool alongside a summary report'' (P1)). 
{\small
\begin{table}[htp]
    \caption{Inter-rater reliability agreement across interviews.}
    \begin{center}
    \begin{tabular}{c c c}
    \toprule
    Interview	& Fleiss' Kappa	 & Interpretation of agreement\\
    \midrule
    1	& 0.42 &Moderate\\
    2	& 0.52 &Moderate\\
    3	& 0.44 &Moderate\\
    4	& 0.73 &Substantial\\
    5	& 0.74 &Substantial\\
    6	& 0.84 &Almost perfect\\
    7	& 0.71 &Substantial\\
    8	& 0.67 &Substantial\\
    9	& 0.62 &Substantial\\
    10	& 0.86 &Almost perfect\\
    11	& 0.70 &Substantial\\
    12* & 0.74 & Substantial\\
    \bottomrule
    \end{tabular}
    \end{center}
    \label{table:interraterInterviews}
    \end{table}%
    }
    
\noindent
\subsubsection{Reflections.}
Our experience surfaced several observations when eliciting and refining goals for LEA in the ML context:

\noindent
\textbf{R1. Abstract stakeholder language:} Several goals were expressed using terms such as ``high harm’’ or ``life-threatening.’’ 
These were context-dependent and interpreted differently across stakeholders, making them difficult to refine into precise, operationalizable goals. 


\noindent
\textbf{R2. Restricted access to verbatim data:} 
Security restrictions meant that only summary notes, rather than full transcripts, were available, reducing context and making it harder to clarify ambiguities or reconcile interpretations during goal refinement without repeated stakeholder involvement.

\noindent
\textbf{R3. Stakeholder scepticism:} 
Some participants expressed reservations about the feasibility of AI in this domain, leading to cautious or underspecified goals (e.g., ``there are times a tool would be helpful, there are other times they would only know through human intuition’’ (P5)). 
  {This required the SE team to carefully distinguish between genuine technical constraints and culturally embedded reservations about replacing human judgment}. 

\noindent
 \textbf{R4. Limited familiarity with AI terminology:} 
Stakeholders rarely used concepts such as prediction'' or human-in-the-loop'' unprompted; these only emerged after explanation and illustrative examples were introduced by the SE team.


\subsection{Data Elicitation}

We introduced data elicitation as a KAOS extension to ground goals in empirical evidence. This process included codebook development and data labelling and later helped bridge stakeholder goals with ML-based operationalization.

\subsubsection{Context.}
 At the start of the project, an initial codebook was available from the predecessor study, created largely by the same team of FCP and FL. In qualitative research, a codebook~\cite{Fereday2006} is a template for organising text for analysis. Each code is identified by a label, a definition of the theme, and a description of how to recognize its occurrence.

In the first stage, 351 behavioural codes from the codebook were applied to dataset D1. This codebook served as a foundation for further coding of subsequent datasets D2--D4. FCP team members elicited new behavioural codes from the data and labelled relevant text segments with those codes. Whenever the codebook was revised (e.g., codes merged or split), earlier labelled datasets were revisited to maintain consistency. 

New behavioural codes were mostly elicited inductively from the data, with some adapted deductively from existing frameworks, integrating bottom-up and top-down perspectives. Whenever the codebook evolved (e.g., merging or splitting codes, introducing new constructs), previously labelled datasets were revisited for consistency. Through this process, the codebook grew to 636 codes, supporting the annotation of 88,976 text segments.

In parallel, the FL team developed a linguistic codebook, starting from 56 linguistic features inherited from prior research and expanding to 173 codes across datasets D1--D4. This effort resulted in 110,922 annotated examples capturing linguistic cues relevant to child sexual abuse investigations. Reliability was rigorously monitored: inter-rater agreement was checked at multiple milestones.
  This reliability work proved essential for later phases of ML model development and feature engineering. 
The SE and ML group members became increasingly involved once stable, reliable annotations were available, using the labelled data to inform object capture and the feasibility of goal operationalization. 

\subsubsection{Reflections.}
Our experience with data elicitation highlighted both challenges and benefits that shaped subsequent iterations:

\noindent
\textbf{R5. Foundation for goal operationalization:} 
The behavioural and linguistic codebooks evolved iteratively alongside goal elaboration. As codes were added, merged, or split, previously labelled data had to be revisited to maintain consistency across datasets.
%

\noindent \textbf{R6. Interdisciplinarity as a strength:}
Close collaboration between FCP experts and SE/ML developers influenced how the system scope evolved. Decisions were required when technically feasible features did not clearly align with stakeholder goals.

\noindent
\textbf{R7. Terminology and granularity gaps:} 
Parts of the inherited codebooks reflected finer-grained distinctions than those used by stakeholders when articulating goals. This mismatch in abstraction and terminology required additional effort during later iterations of goal elaboration and object capture.
      

\subsection{Object Capture}

KAOS object capture focuses on identifying entities, relationships, and attributes that underpin the domain and give semantics to goals and operations. In our ML-enabled setting, however, many conceptual objects could not be fully specified from stakeholder interviews alone. The abstractness of terms such as ``high harm” or ``suspect engagement” meant that objects required empirical grounding in the elicited data to become operationalizable.

\subsubsection{Context.}
\begin{figure}
\center
\includegraphics[width=0.35\textwidth]{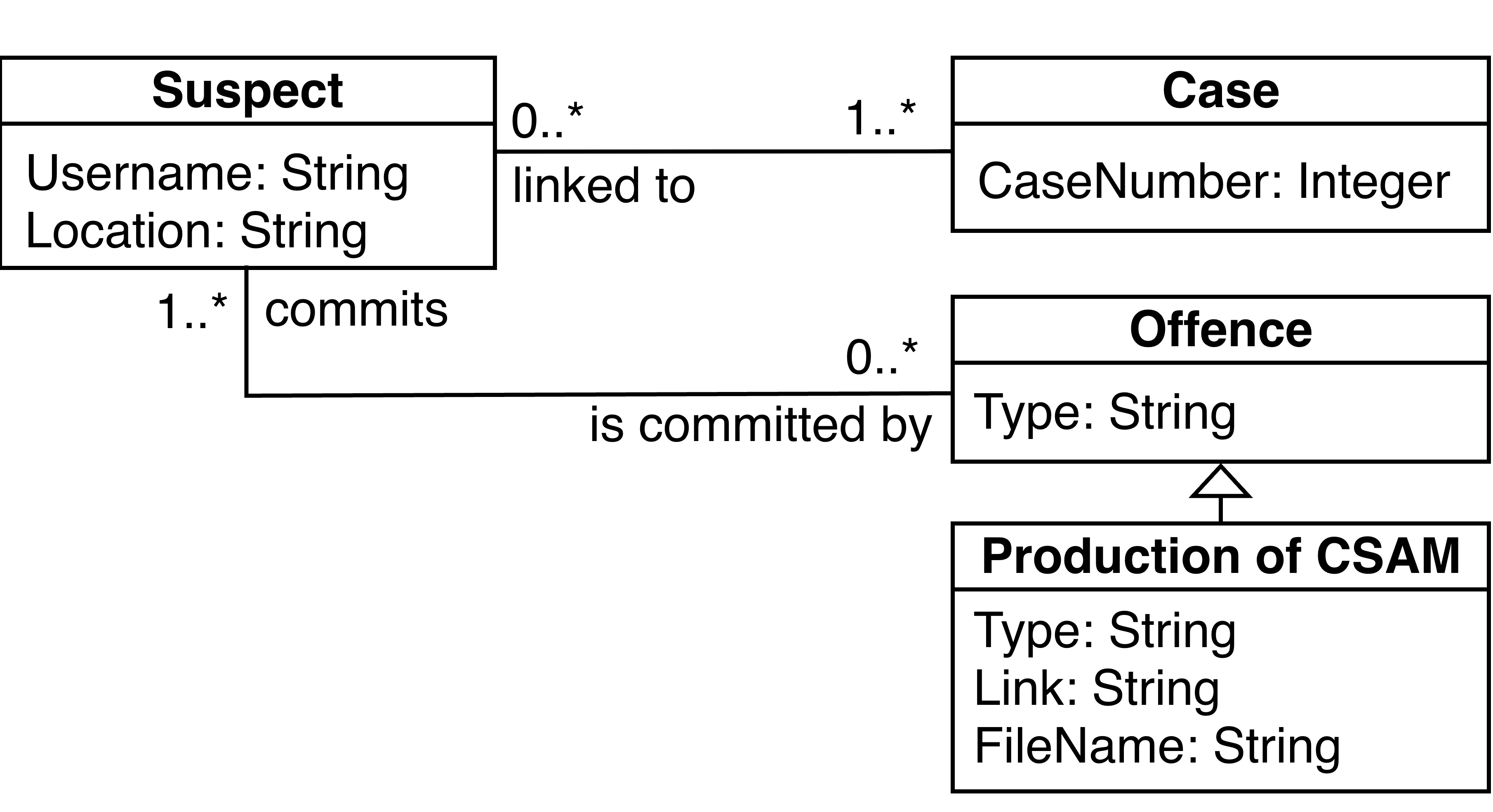}
\caption{Object model fragment for the ML-based system.}
\label{figure:object_model}
\end{figure}

From the elicited goals, an initial set of  objects was defined. These objects captured key investigative concepts and their relationships within the domain. 
For example, in a law enforcement setting, relevant entities  included \texttt{Suspect}, \texttt{Case}, \texttt{Offence} and Production of Child Sexual Abuse Material (\texttt{Production of CSAM}) as one offence category, along with associations, such as ``suspect linked-to case'',  ``suspect commits offence'' and ``offence is-committed-by suspect'' (see Fig.~\ref{figure:object_model}).

Because these objects were intended to provide semantics for ML tasks, they also needed to be grounded in the data available for training and evaluation.  Mapping between concepts in goals and those that arose in the data (codes)  became a central activity in object capture. 
To this end, a focus group was held  with four researchers involved in the behavioural and linguistic coding of D1 and D2, where initial themes were developed and then led to the construction a provisional mapping between objects in the goal model and behavioural codes. Linguistic codes were sometimes composed to approximate a single object.  This mapping was iteratively refined through cross-team moderation (forensic and clinical psychology, forensic linguistics, and software engineering). An example of this is given below where the attribute ``Category'' of ``Production of CSAM'' entity is mapped to two codes conjoined.

{\small\begin{alltt}
Entity: Production of CSAM
	Attribute  Category  \guillemotleft groundedBy \guillemotright ImageContainsChild 
					 								\texttoplus{} ImageDepictsSexualActivity
\end{alltt}}

In some cases, objects extracted from goals had no associated label in the data, prompting augmentation of the codebook in the first instance,  requests for additional datasets, or refinement of the goal model when neither was possible. Conversely, some codes proved too broad and had to be split into finer-grained subcodes before alignment was possible, to  align with objects from the goal model. The final mapping linked 106 behavioural codes (17\% of total) and combinations of 59 linguistic codes (34\% of total) to 30 objects referenced in the goal model.

\subsubsection{Reflections.}
The following insights emerged at this stage:
%
 
 \noindent
 \textbf{R8. Ambiguity and refinement options:}  
 Many candidate objects were expressed in high-level or subjective terms, risking abstraction without clear grounding. Examining how these concepts appeared in the labelled data helped disambiguate them and revealed alternative refinement paths and design choices.

\noindent
\textbf{R9. Focusing labelling effort:} Early iterations distributed labelling effort across a large number of codes. Once object--code mappings became clearer, later iterations allowed labelling to focus on features directly relevant to operationalizable goals.

\noindent
\textbf{R10. Identifying scope and feasibility:} 
Mapping objects to available codes exposed which stakeholder goals could realistically be supported by the data. Goals without corresponding codes either required codebook augmentation or were marked as infeasible, enabling early scoping decisions.

\subsection{Responsibility Assignment to ML Agents}

In KAOS, responsibility assignment requires identifying which agent (human, software, or environmental) is responsible for satisfying a leaf goal. Traditionally, this involves characterizing the agent’s interface (the variables it can monitor and control), its transition system (mapping sequences of monitored and controlled variables to next states), and the responsibility relation (mapping the agent to the goals it must satisfy).

\subsubsection{Context}
For ML agents, the definition of an interface diverged from standard KAOS practice. Interfaces were reconstructed indirectly from what could be elicited and encoded as training labels, together with the available input modality. This introduced uncertainty at two levels: first, whether the relevant phenomenon was sufficiently observable in the input, and second, whether the ML agent could recognize it reliably. Consequently, monitorability was not absolute but conditional, and realizability depended not only on system/environment assumptions, as in classical KAOS, but also on the adequacy of data and the learnability of the target.

Assignment decisions in our project were grounded in artefacts from earlier phases, particularly data elicitation and object capture. The availability of labelled data constrained which ML agents could be considered: if no suitable labels existed, a goal could not be assigned to an ML agent. The object model clarified how text segments (our primary input) related to domain entities and target features. Goals whose satisfaction conditions could be approximated by observable text features, and for which labelled data existed, were provisionally assigned to ML agents. (See Fig.~\ref{figure:KAOS}) Goals requiring reasoning over unobservable properties, lacking data, or demanding high-stakes precision (e.g., legal thresholds) were deferred to human investigators or deterministic software.  
Responsibility assignments were treated as provisional and iteratively refined as models were developed and tested. Where multiple ML agents might satisfy a requirement, this was captured in KAOS as an OR-refinement. Poor model performance or missing features prompted revisiting allocations and redistributing responsibilities between ML, deterministic software, and human agents.

\subsubsection{Reflections.}
The following insights emerged:

\noindent
  \textbf{R11. Indirect mapping of features:} Stakeholder goals could not always be linked directly to observable features. Monitorability was therefore conditional, requiring ML agents to infer internal representations, which reduced transparency.

  \noindent  \textbf{R12. Data-constrained realizability:} Goals with insufficient examples in the available data could not be realized by ML agents and were either assigned to humans or excluded from scope. The variability of natural language inputs further complicated learnability and limited reliable automation.

\noindent    \textbf{R13. Iterative reassignment:} Responsibility assignments were provisional. Weak model performance required revisiting assignments and redistributing responsibilities across ML, deterministic software, and human agents.

\noindent     \textbf{R14. Cross-phase dependence:} Responsibility assignment depended directly on outcomes of data elicitation and object capture. Label availability and entity--feature mappings shaped realizability, making the KAOS phases mutually constraining and iterative.

\subsection{Goal Operationalization with ML Models}
In KAOS, leaf goals are operationalized into requirements that constrain the operations of assigned agents, typically through pre-, trigger-, and post-conditions. This section focuses exclusively on the operationalization of goals assigned to \emph{ML agents}, for which these classical assumptions do not hold.


\subsubsection{Context}
For ML agents, an ``operation'' corresponds to a model inference step that consumes input data and produces a probabilistic output rather than a deterministic state transition. Consequently, goal satisfaction cannot be verified through crisp post-conditions but must be assessed empirically using labelled data. In our project, ML operational requirements were realized as probabilistic binary classification tasks: each model received a text segment as input and returned the probability that a target investigative feature was present.
Two families of ML agents were instantiated: fine-tuned BERT models~\cite{BERT}, which produced confidence scores for specific features, and probabilistic inductive logic programming (ILP) learners~\cite{LAWMARK2023CILP}, introduced to address requirements with sparse labelled data.

To express ML operationalization, operation specifications were reformulated to include $(i)$ an \emph{input condition} describing the available data structure, $(ii)$ an \emph{inference specification} defining the probabilistic mapping (e.g., $M(x) \mapsto P(\textit{feature}\mid x) \in [0,1]$), and $(iii)$ a \emph{performance guarantee} expressed as quantitative thresholds on evaluation metrics (e.g., precision $\geq 0.75$, recall $\geq 0.85$), serving as probabilistic analogues of classical post-conditions.


Out of 30 leaf goals, 15 were assigned to ML agents and operationalized in this manner. The prototype included 15 BERT models and 10 ILP learners. Table~\ref{table:evaluation2} presents representative results for three BERT and three ILP models, illustrating performance under varying amounts of labelled data. Across these models, precision ranged from 0.62--0.92 (BERT) and 0.68--0.83 (ILP), while recall ranged from 0.78--1.00 and 0.70--1.00, respectively. When training data were sparse (72 labelled examples), replacing BERT with an ILP learner improved precision to 0.83, recall to 1.00, and F1 to 0.91.

{Out of 30 leaf goals, 15 were assigned to ML models (BERT and some ILP), 13 were assigned to non-ML, traditional software and two were assigned to LEA investigators.}
Leaf goals were then operationalized with such probabilistic specifications and evaluated empirically. In total, 15 BERT models and 10 ILP learners were included in the prototype of the system. In Table~\ref{table:evaluation2}, we present an evaluation of three BERT and three ILP  models to illustrate how model performance compares when trained with a large versus a small number of labelled (positive) examples. Recall was prioritized over accuracy because false negatives were considered riskier than false positives in the investigative setting; this trade-off was captured as a soft goal in the KAOS model. Operationalization also revealed dependencies on data structure: long text segments labelled with behavioural codes hindered learning and required relabelling into smaller units, while aggregating subcodes into higher-level codes increased the number of training examples and improved model performance.


 {\small
\begin{table}[htp]
\caption{Example of model performance comparison.
}
\begin{center}
\setlength\tabcolsep{1.8pt}
\begin{tabular}{c | c | c c c | c c c }
\toprule

\multirow{2}{*}{Model} & \multirow{2}{*}{\# labelled examples} 
& \multicolumn{3}{c|}{BERT models} & \multicolumn{3}{c}{ILP learners} \\
\cline{3-8}
 & & precision & recall & F1 & precision & recall & F1 \\
\midrule

1 & 957 & 0.92 &	0.94 &	0.93 & 0.77 &	0.75 & 0.76 	\\
2 & 664 & 0.80 &	0.78 & 0.79 &	0.68 &	0.70 & 0.69 	 \\
3 & \textbf{72} & 0.62	& 1.00 & 0.77 & \textbf{0.83} & 1.00 & \textbf{0.91} 	 \\
\bottomrule
\end{tabular}
\end{center}
\label{table:evaluation2}
\end{table}
}

\subsubsection{Reflections.}
Three insights emerged:

\noindent
\textbf{R15. Dual sources of probability:} 
Unlike deterministic software agents, ML agents satisfy goals only probabilistically. Uncertainty arises both from the partial observability of features in text and from model prediction accuracy. As a result, goal satisfaction had to be expressed through prediction-time probabilities and empirical performance metrics rather than crisp post-conditions.
%

\noindent
\textbf{R16. Sensitivity on data structure:} Model performance was strongly shaped by data quality and granularity. Relabelling long text segments into smaller units improved learning, while sparse datasets motivated the use of probabilistic ILP learners that could exploit structural knowledge.

\noindent \textbf{R17. Feedback into earlier phases:} 
Failure to meet required performance levels (e.g., insufficient recall) triggered revisions in earlier phases, including reassignment of responsibilities, refinement of monitored objects, or augmentation and restructuring of labelled data. Recall was treated as a soft goal to reflect the higher risk of false negatives in the investigative context.

\section{Lessons Learned and Directions for KAOS}
\label{section:lessons_learned}

Our experience indicates that KAOS provides a valuable foundation for structuring requirements in ML-enabled systems for LEAs, but also exposes systematic gaps in this setting. \emph{Classical KAOS} implicitly assumes that requirements can be refined without explicit consideration of data availability, that domain objects can be specified independently of empirical datasets, and that operationalizations can be deterministically derived and assigned to agents via well-defined interfaces.
In our project, progress across all phases depended on making \emph{explicit links between goals and empirical evidence in data}. Data availability and quality constrained goal refinement, feasible operationalizations, and responsibility assignment, revealing the need to treat \emph{data elicitation} as a first-class activity in goal modelling---currently addressed only implicitly in KAOS.
In the following, we synthesize lessons learned, grounded in reflections R1-R17 reported in Section~\ref{section:applying_KAOS}, and propose targeted extensions to KAOS that make these data dependencies explicit.
%

\subsection{Goal Elaboration}
\noindent
\textit{Lesson Learned:} 
KAOS provided a useful structure for eliciting and refining stakeholder goals. 
In the ML/LEA setting, refinement often stalled when abstract or ambiguous goals 
(e.g., ``prevent high harm”) could not be grounded in empirical data (R1). 
Progress required deliberate effort to: $(i)$ translate stakeholder language into concepts and signals 
represented in available datasets; $(ii)$ work around confidentiality restrictions 
that limited access to verbatim case material (R2); and $(iii)$ introduce ML concepts in forms 
that investigators could meaningfully reason about (R4).

A key difficulty was choosing the \emph{right level of refinement}. Goals refined too narrowly 
(e.g., into very fine-grained behavioural categories) associated with labels with too few positive examples 
to train reliable models. Goals kept too broad resulted in large, mixed datasets that were difficult 
for models to learn accurately and for investigators to trust. This ``granularity gap’’ (R1, R3) 
meant that goal refinement had to be guided by feedback from data elicitation (label coverage, 
class balance) and from early operationalization experiments (model learnability and performance).

\noindent
\textit{Possible direction:} Extend KAOS goal elaboration to explicitly link leaf goals to data artefacts (e.g., datasets, labelled codes, labelling requirements). Making these dependencies explicit would help engineers identify unrealizable goals early and trigger refinement, re-scoping, or targeted data collection.


\subsection{Data Elicitation}
\noindent

\noindent
\textit{Lesson Learned:} 
In our project, \emph{all KAOS activities were constrained by data availability and quality}: goal elaboration, object capture, and responsibility assignment could not proceed independently of reliable, sufficiently granular, and consistently labelled data. As observed during iterative codebook evolution (R5), changes to behavioural and linguistic codes repeatedly required revisiting earlier annotations, with direct impact on downstream operationalization. Consequently, early iterations adopted \emph{wide, exploratory coding} to assess which goals the data could realistically support, to refine goals by exposing how stakeholder concepts were expressed in practice, and to guide dataset structuring—particularly where inherited codebooks introduced abstraction mismatches with stakeholder terminology (R7).

Establishing a \emph{gold standard codebook} became essential once this exploratory phase stabilized. It anchored annotation and inter-rater reliability, clarified what could be consistently coded, and surfaced data gaps and ambiguities. Beyond quality control, the gold standard provided a shared and objective basis for prioritizing data augmentation and early ML feasibility testing, supporting coordination across disciplinary boundaries (R6).
As high-level requirements stabilized, elicitation shifted to \emph{goal-driven labelling}, prioritizing codes traceable to stakeholder goals and suitable as ML features. This reduced rework from evolving abstractions (R5), narrowed terminology gaps (R7), and enabled earlier feasibility checks and testing of candidate ML models.

%

\noindent
\textit{Possible direction:} 
Introduce a KAOS \emph{data model} that captures datasets, label sets, and quality indicators (e.g., coverage, balance), and links them to the goal model via traceability relations. Such a model could also connect to the KAOS \emph{agent model}, enabling reasoning about which agents (human or software) can fulfill responsibilities given the available data. This would allow systematic triggering of actions to: $(i)$ refine object--label mappings; $(ii)$ augment or relabel data; or $(iii)$ reassign goal responsibility when data constraints prevent goal satisfaction.


\subsection{Object Capture}
\noindent
\textit{Lesson Learned:} 
In ML-enabled systems, KAOS object capture cannot be treated as a one-off conceptual modelling step. Objects only become meaningful once their empirical grounding is established, and this grounding must be revisited as goals, codes, and datasets evolve. As reflected in R8--R10, object capture therefore functions as an iterative negotiation between stakeholder abstractions, available data, and feasibility constraints. More generally, grounding objects in data enables early reasoning about refinement choices, labelling effort, and goal feasibility, allowing object capture to directly inform scoping decisions for downstream ML development rather than merely documenting domain concepts.

\noindent
\textit{Possible direction:} Extend KAOS object capture to treat data artefacts as first-class modelling elements. Making the link between conceptual objects and empirical evidence explicit would support reasoning about granularity and feasibility trade-offs and enable earlier, data-aware scoping decisions in ML-enabled systems.

\subsection{Agent Assignment}
\noindent
\textit{Lesson Learned:}  
In ML-enabled systems, classical KAOS assumptions about deterministic responsibility assignment break down. As reflected in R11--R14, assigning leaf goals to ML agents proved inherently \emph{provisional and data-dependent}. Goals could only be allocated to ML agents when the relevant phenomena were observable in available inputs, supported by sufficient labelled data, and learnable with acceptable predictive performance.

More generally, monitorability became conditional, realizability probabilistic, and responsibility assignment an iterative process. Reassignment triggered by poor model performance exposed data gaps, revealed goals that were not practically achievable, and informed early scoping and architectural decisions. Responsibility assignment thus functioned as a recurring \emph{feasibility check} tightly coupling data, model behaviour, and system design.


\noindent
\textit{Possible direction:} Extend KAOS responsibility assignment with an explicit notion of \emph{data realizability}, expressing the feasibility of assigning goals to ML agents based on data availability, labelling quality, and learnability, and enabling reasoning about partial or probabilistic satisfaction under data uncertainty.

\subsection{Goal Operationalization}
\noindent
\textit{Lesson Learned:} 
Classical KAOS operationalization assumes deterministic operations whose correct execution guarantees goal satisfaction. Our experience with ML-enabled agents shows that this assumption does not hold: operations correspond to probabilistic inferences, and goal satisfaction can only be demonstrated empirically ({R15}). Moreover, operational success depends critically on data availability, labelling granularity, and structural properties of the dataset ({R16}), making operationalization inseparable from data elicitation. When empirical performance was insufficient, this failure systematically propagated back into earlier KAOS phases, requiring revisions to responsibility assignment, monitored objects, or data artefacts ({R17}).

\noindent
\textit{Possible direction:} To address these limitations, KAOS operationalization could be extended with an explicit notion of \emph{partial satisfaction} for ML-enabled operations. While goals would remain expressed in stakeholder terms, their operationalization would acknowledge probabilistic satisfaction, evidenced through empirical metrics such as recall or F1 on representative datasets. This extension would make data dependence and feedback loops to earlier modelling decisions  first-class concerns, while preserving the ML-agnostic nature of the goal model.
\section{Threats to Validity}
\label{section:threats}

\noindent\textbf{Internal validity.}  
Two members of the research team with expertise in forensic psychology conducted the interviews with LEA participants. To mitigate potential bias stemming from the interviewers' disciplinary backgrounds, the team collectively prepared and refined both the main and follow-up questions in advance.  

To reduce interpretive bias during analysis, two researchers independently performed thematic coding of the interview notes. Discrepancies were discussed and resolved in consultation with an academic experienced in requirements engineering and process automation in law enforcement.  

Another threat arose from the partial ambiguity of certain interview notes. For confidentiality reasons, sensitive details were redacted and some notes were generalised. We sought to preserve the intended meaning of participant contributions and used the available material as representative indicators of our key findings.  

Finally, the study was situated in the sensitive context of law enforcement, where requirements processes are shaped by confidentiality constraints, limited access to verbatim data, and the high stakes of investigative decision-making. These factors influenced both how KAOS could be applied (e.g., reliance on summary notes rather than transcripts) and how goals were prioritised (e.g., recall was treated as more critical than overall accuracy due to the risks of missing a relevant suspect).  

\noindent \textbf{External validity.}  
Our interviews involved three LEA teams working on the crime type targeted by the ML-enabled system, each with a distinct investigative focus. Interviewing across these teams allowed us to capture a broader set of requirements for system design.  
Nonetheless, our findings may not extend to investigators tackling other crime types, working in different units, or operating in other LEAs. We caution against assuming direct generalizability beyond the studied setting. As an experience paper, our aim is not to propose a universal method, but to show how KAOS can be applied in practice and to highlight extensions worth further exploration by the requirements engineering community.  

That said, we believe the work offers transferable insights for designing ML-assisted decision-making tools in law enforcement and other high-stakes domains. In particular, linking goals to labelled data, reasoning about probabilistic rather than deterministic agent behavior, and treating empirical performance as part of operationalization are challenges common to many ML-enabled systems, from healthcare to finance. What varies across domains are the soft goals prioritised and the organizational constraints on data elicitation.

\section{Addressing Ethical Considerations}
\label{section:ethics}

The use of ML in decision-making raises ethical concerns around transparency, accountability, privacy, and potential unfairness~\cite{Osasona2024Reviewing}, amplified in law-enforcement contexts. In our project, these concerns were addressed procedurally (e.g., approvals and oversight) and technically (e.g., explanation mechanisms and evaluation), and can be made explicit within KAOS and its extensions.

Within KAOS, transparency was represented as a soft goal, refined into operational requirements such as a software agent generating explanations alongside predictions and retaining decision traces prior to action. Accountability and oversight were captured as assumptions on the \emph{Human Investigator} agent: predictions were treated as advisory, validation was mandatory, and action only followed human sign-off, ensuring that decision authority remained with investigators. Privacy and data governance were captured as domain assumptions and constraints linked to a \emph{Data Protection Officer} agent, responsible for approving datasets and their use prior to training and deployment, thereby enforcing governance gates. Fairness was treated as a soft goal: although we did not observe explicit unfairness in manual labelling, the risk of bias was acknowledged and mitigated by assigning investigators responsibility for verifying predictions and flagging biased or unfair outcomes. Such feedback was incorporated into model retraining to reduce the likelihood of repeated bias. Robustness and quality were supported by training models on high-quality, manually labelled datasets curated across multiple platforms by our interdisciplinary team. In our domain, recall was prioritised over precision, reflecting the risks associated with missing a relevant suspect.

{The research was granted full ethical approval by the Science, Technology, Engineering and Mathematics Ethical Review 
Committee at the University of Birmingham, }
which served as the lead institution. Ethics committees at partner institutions accepted this process. The team adhered to the universities’ Codes of Practice for Research, and researchers working on the datasets held appropriate security clearances with the funders.

Finally, our approach is consistent with public-sector guidance on ethical AI~\cite{Ethics2023} and with the \emph{Covenant for Using Artificial Intelligence in Policing}~\cite{NPCC2023}, by making oversight and accountability explicit, enforcing governance gates, retaining decision logs, and acknowledging the probabilistic nature of ML outputs. While confidentiality prevents the release of datasets and trained models, the KAOS-based linkage between goals, data, objects, agents, and operations provides a transferable template for designing, auditing, and evolving ML-enabled systems in other sensitive domains.

\section{Conclusion}
\label{section:conclusion}

This paper reported our experience applying KAOS to an ML-enabled decision support system in law enforcement. While KAOS effectively structured stakeholder objectives, trade-offs, and responsibility assignment, our study exposed systematic gaps when classical concepts met ML’s probabilistic, data-driven nature. We showed that tightly coupling goal modelling with bottom-up data elicitation and labelling strengthened goal refinement, surfaced missing domain concepts, revealed dataset limitations affecting fairness and explainability, and improved ML interpretability and stakeholder confidence. Future work will formalize these extensions---data realizability, empirical operationalization, and data-aware elicitation---to support systematic application in other ML-enabled systems.

\section*{Acknowledgement}
We would like to acknowledge financial support from the University of Birmingham and Aston University's EPSRC Impact Acceleration Award (IAA) fund, and University of Bath's EPSRC and ESRC IAA funds. We also thank Mr Fahim Ahmed for his contributions to the later stages of the project, particularly in the design and evaluation of the system’s user interface which are outside the scope of this paper.

\newpage

\bibliographystyle{ACM-Reference-Format}
\bibliography{mybib}

%
%
%
%
%
%
%
%

\end{document}